# Hybrid modes in multilayer/antenna architecture set sideband-selective molecular Raman scattering

Giuseppina Simone§

**Abstract**:  In the realm of healthcare, label-free sensing is a vital component for various applications. Micro-photonic technology presents a promising avenue to pursue pivotal goals. With the use of this technology, healthcare professionals and researchers can harness the power of label-free sensing to develop effective diagnostic and therapeutic tools. The integration of label-free sensing and micro-photonic technology can lead to revolutionary advancements in the healthcare industry. Here, a hybrid multilayer, whose surface is decorated with micrometric features has been investigated to evaluate the application in this direction. Following the characterization of the plasmonic resonance, and upon demonstrating the surface plasmon polariton (SPP)/antenna mode hybridization, the optomechanical behavior of the system has been investigated. The hybrid system has a higher sensitivity due to its high-quality factor $Q$ when compared to standard systems. The shift of the frequency is studied for a red dye, at different excitation wavelengths as well as in an aqueous environment. To validate the increased sensitivity, I conducted an analysis using bovine serum albumin (BSA). This protein is water-soluble and has an infrared absorption band (amide I), and it is also active in the Raman region. Moreover, it can consistently bind to Ag features. Through an image-based analysis, the surface-enhanced Raman scattering pattern of the BSA was recorded. The proposed sensing method is innovative and opens new perspectives on sensitive methods for biomolecule detection. Simultaneously, it shows promising results to exploit the optical resonance shift as a basic sensing approach for probing molecular patterns.

## 1. Introduction

High-sensitive detection of molecules is crucial for several applications, such as diagnosis, drug discovery as well as security screening. In the last decades, optical detection has attracted great interest in this field thanks to its high-quality factor $Q$ and high sensitivity down to the molecular scale [1]. Moreover, its label-free approach is very beneficial for many reasons. There has been a significant focus on examining individual molecules through photonic analysis [1,2]. A resonance shift of about $10^{-10}$ with systems having $Q=10^6$ in an aqueous environment can be achieved, which is an important stepstone in molecular analysis but still far from the sensitivity for detecting a single binding event [3,4]. Optical cavities have been studied for molecular analysis [5,6]. The sensing mechanism of the cavities critically depends on the optical quality factor; molecule/high $Q$ cavity coupling perturbs the cavity mode leading to the shift of the resonance and a change of the cavity transmission as well [7].

The high $Q$- factor derives from a linewidth that is narrower than the vibrational frequency of the molecular species and imposes the crucial hierarchy in a resolved side-band regime, where the mechanical frequency exceeds the optical linewidth as a fundamental condition. However, the higher sensitivity related to a high $Q$- factor represents a limitation because of the losses associated with the resonance of the system, which means that the linewidth is much larger than the frequency with a consequent conspicuous decay rate [8–10]. Therefore, the losses limit the performance of the figure-of-merit of the resonant system. To reduce the linewidth of the plasmonic resonance, several strategies based on plasmonic hybridization have been studied [11–13] Hybrid plasmonic systems including dielectric and metal layers combined with nanometric features allow for design tuning and flexibility, which resulted in a variety of resonance line shapes and cavity modes [14,15]. Although the hybridization between the SPPs and the cavity requires a higher level of complexity to be understood [16], it offers several advantages. The optical antenna offers excellent confinement of the electric field for large coupling to the molecular dipole [12,13,17–19], at the price of a significant reduction of the effective detection area.

The wave that is driven inside the cavity by the photons leads to a radiation pressure that couples with the mechanical motion of the oscillator and enables the quantum control of the mechanical motion [20–22]. Furthermore, because of the blue detuning of laser wavelength to the cavity resonance, the optical wave enhances the mechanical motion above the threshold of oscillation, resulting in highly coherent optomechanical oscillation with a narrow mechanical linewidth [23–25]. The optical wave inside the cavity induces mechanical rigidity and simultaneously shifts the system to an optomechanical frequency that depends on the laser/cavity detuning. This permits that the perturbation of the cavity resonance caused by the binding of analytes to the cavity is transduced in a conspicuous shift of the mechanical oscillator frequency [26]. As the efficiency of the transduction mechanism strengthens resonance sensing, the optically induced frequency shift of the optomechanical oscillation has been used in the last decade to transduce and amplify the molecular binding signal.

In this context, here a hybrid system consisting of a multilayer integrating plasmonic nanometric features generated by the etching of the silicon and showing an augment of $Q$-factor is studied. The enhancement has been attributed to a localized acoustic vibration of the roughness features [27–31]. The system indeed is made of parallel piled layers, alternating high-quality optical mirrors with good reflectivity and dielectric layers. The mirror heights are designed to be significantly larger than the surface plasmon polariton penetration depth in the air so that the losses are reduced. The preliminary analysis aimed at characterizing the hybridization strength of the SPP/multilayer system [32,33]. First, the plasmonic behavior of the cavity is described and the $Q$-factor is determined; then, the numerical and experimental results are compared and discussed, to provide an overview of the electric and magnetic field and to analyze the hybridization between the localized SPPs and the multilayer mode [34–38]. The analysis surges to higher interest when molecules are coupled to the hybrid system, due to the enhancement of the optomechanical coupling between the localized surface plasmon resonance and the molecular vibrational mode [39,40]. Indeed, the detection of biomolecules still represents an open challenge in the sensor field. There is a clear interest in operating simple protocols, which can guarantee both high sensitivity and repeatability at the same time. The proteins belong to the family of biomolecules, and they are, perhaps, the most interesting because of the fundamental role they play in diagnosis and prognosis. For increasing the capacity of molecular sensing, it is suggested that either high sample concentrations or very highly sensitive optical oscillators that can detect fluctuations are needed. To date, the bovine serum albumin (BSA) has been tested, and a strong coupling has been demonstrated for this complex system. The present configuration provides a notable advantage in achieving light-matter coupling. Specifically, the interaction takes place between the single plasmon mode of the hybrid cavity and the molecular mode. This coupling of light and matter is of utmost importance in the fields of optics and photonics, as it has the potential to revolutionize the design of photonic devices. By exploiting this coupling, it is possible to achieve unprecedented control over light and matter interactions, leading to improved performance and efficiency of the devices. Therefore, this achievement represents a significant step forward in the development of state-of-the-art photonic devices. The mechanical resonators of a system, composed of a multilayer and molecules, operate at significantly different typical frequencies, with the multilayer operating at several MHz and the molecules at a few tens of GHz. The system's ability to interact with and probe its surrounding environment depends on these frequencies. In the form of surface plasmon polaritons (SPPs), light hybridizes with the cavity and couples with the molecule, inducing plasmon-exciton coupling. This reveals molecular vibrational modes and a region rich in molecular transition [41].

In conclusion, based on these considerations, the cavity/resonator hybridization is expected to enhance the sensitivity of the molecular analysis. This is proved by the experimental investigation, which aimed to demonstrate that the analysis of the hybrid multilayer/BSA transmitted signal allows the detection of the Raman signal correlated to the molecular vibration of the system. Indeed, coherently with standard Raman methods, the spectrum displays the principal peaks of the proteins.

## 2. Results and Discussion

### 2.1. Characterizing the multilayer hybrid system

The scheme of the experimental system under investigation is depicted in **Figure 1a**. It comprises a groove designed for hosting the sample. The primary function of this groove is to provide fluid dynamic stability to the sample. To generate the optical mode, the active section consists of alternating layers of dielectrics and noble metals. The morphology of the surface is characterized by having micrometric features resulting both from etching residues and the extreme anisotropy of a process that has been realized by fast alternated etching and passivation steps cover the surface. The micrometric features work like antennae are schematized as a rectangle and the layout is shown in **Figure**

**1b**. The rectangle has an average diameter of 1.4±0.3 µm and an average height of 30±5 µm, while the antenna-to-antenna distance (gap) is *b*= 1±0.3 µm.

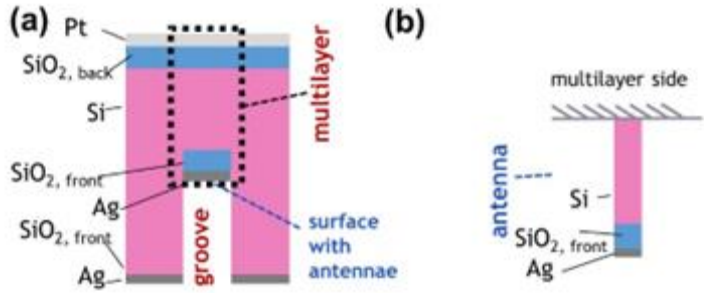

**Figure 1.** Design of the hybrid multilayer. (a) Schematic cross-section of the multilayer and etched groove. Note that the diagram is not to scale (b) Not-to-scale schematic cross-section of the antenna.

It is hypothesized that the micrometric features present in the multilayer structure enhance the generation of surface plasmon polaritons (SPPs) at the interface with the external medium, while simultaneously bestowing upon the multilayer a hybrid mode. The basis for this hypothesis is derived from the outcomes presented in reference [42].

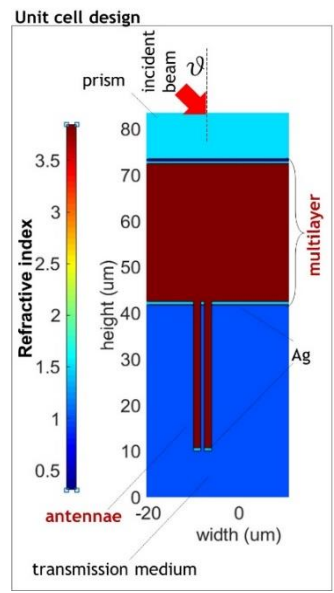

**Scheme 1.** Unit cell design consists of a cross-section of the multilayer and antennae schematized by a rectangular geometry. Color bar: refractive index.

The reflectance and the electromagnetic field distribution were studied for studying the optical mode. The understanding of the role and the impact of the antennae on the plasmonic behavior was studied referring to the geometry represented in **Scheme 1**. In this schematization, the groove has been neglected after having proved to not affect the optical response of the system. Figure 2a displays the result of the simulated reflection spectrum of the plain and planar *sans*-antennae multilayer at different angles of incidence at TM polarization. The analysis reveals that the energy of photons remains unaffected by the angle of incidence; in turn, the antennae of the system result in a hybrid mode, which is characterized by a two-mode crossing at $\vartheta_{res} \approx 43.48$ deg, as depicted in Figure 2b.

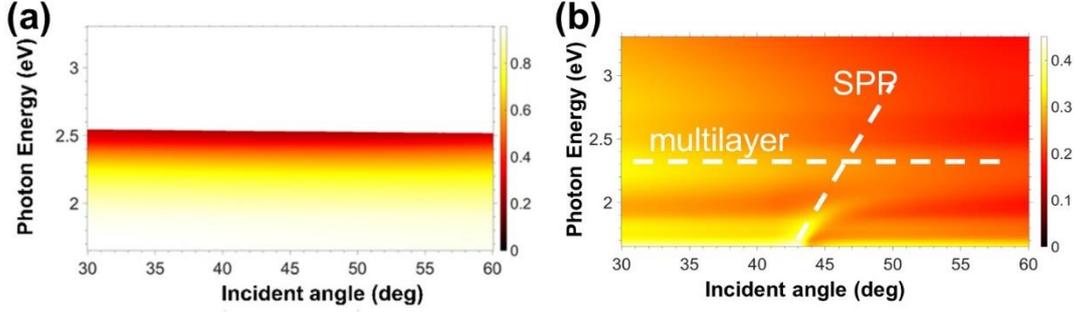

**Figure 2.** Dispersion of the cavity/antenna system. Simulated reflectivity as a function of the photon energy and incident angle for (a) the plain planar multilayer. The white space depends on the conversion of the dispersion plots from energy-momentum space to energy-angle; (b) the multilayer including the antennae. Color bar: reflectivity.

The mode hybridization has been further studied by observing the distribution of the electromagnetic field. Figure 3a displays the distribution of the electric and magnetic fields, whereas Figure 3b provides a closer view of the micrometric structures. The transverse electric modes are denoted as antisymmetric or dark ($E_x$) and symmetric or bright ($E_z$) modes. The dark mode is restricted to the surface, and it does not couple the radiative continuum and is therefore characterized by zero radiative losses. In turn, the bright modes display the the coalesce of the electromagnetic line field at the antennae. Indeed, a thorough analysis of both panels in the figure reveals that when the SPP is excited at an angle of 43.48 deg, the electric field components ($E_x$ and $E_z$) are enhanced in correspondence to the gap (b) between the two antennae and at the top of the antennae. The strength of the electric field's components is primarily influenced by the distance between antennae. While the height and diameter of the antenna do have an impact substantial as the antenna-to-antenna distance. While one can observe that the effect of the antennae on the magnetic field (H) is negligible, Figure 3c highlights that the $E_x$ component experiences a steeper decline compared to the $E_z$ component as the gap between antennas widens. Altogether, the excitation of surface plasmon polaritons generates a local electromagnetic field enhancement in the gap between the antennae, which demonstrates near-field plasmon coupling of light to plasmons. The enhancement of the $E_z$ on the top of the antennae justifies the transmission spectrum trend observed in Figure 3, indeed, the gap is such to allow the plasmonic coupling and enhancement. Moreover, due to the correlation between the enhancement of the electric field and the enrichment of the electromagnetic state density [43], the enhancement of the $E_z$ provides evidence of the hybridization of the cavity with the antennae [44].

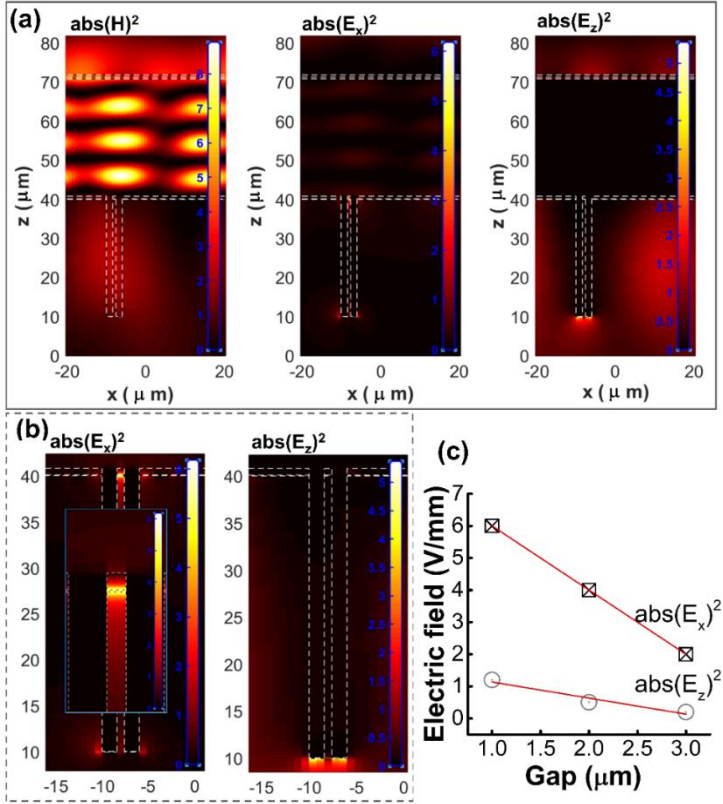

**Figure 3.** Electric ($E_z$ and $E_x$) and magnetic field (H) numerical results from Matlab modeling. The results have been modeled at the incident angle $\vartheta_{res}$ = 43.48 deg and TM polarization. (a) The top panels show the overview of the field components. (b) Zoom at the antenna-to-antenna gap of th electric field components. (c) Electric field maximal intensity at the surface versus the gap size between the antennae. $R^2$=1 for $|E_x|^2$, $R^2$=0.94 for $|E_z|^2$. Color bar: field intensity.

The surface plasmon polariton modes were studied upon illuminating the structure with a transverse magnetic or *p*-polarized beam ($\lambda_{ex}$=650 nm, power 5 mW) in a Kretschmann configuration shown in Figure 4a and well detailed in the Experimental section. To take measurements, the sample was set up with the prism in two different configurations based on the relative orientation of the magnetic field vector $\vec{H}$ and the groove direction (indicated by the arrow in Figure 4b). These configurations were classified as transversal TM1 and parallel TM2. The angular interrogation reported in Figure 4c shows that the TM1 configuration has a resonance at an incident angle of $\vartheta_{res}$ = 43.48 deg and a back-bending of the minimum. The dip of the TM2 configuration is less intense as well as less broad than the TM1 one, while both the curves show a dip at $\vartheta \approx 40$ deg.

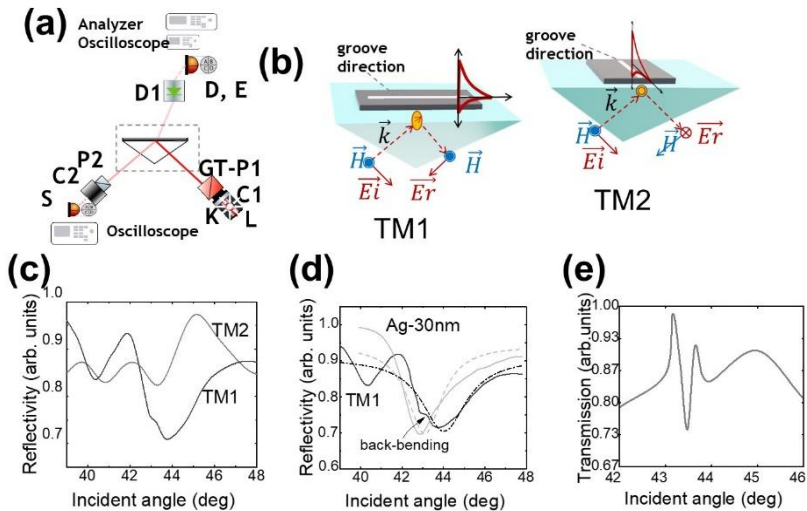

**Figure 4.** Hybridization of the cavity/antenna system. (a) Dual setup including the module of reflectance and transmission (not to scale). L: laser, Ci: Condensator, Pi: Polarizer, K: Chopper, D: Photodiode and oscilloscope, E: Capacitor, D1: Photodetector, S: Spectrometer. (b) Arrangements of the sample relative to the groove orientation. $E, \vec{H}, k$ refers to the electric and magnetic component of the incident (i) and reflected (r) fields, and to the wavevector. (c) Relative reflectivity according to the incident angle and the two configurations TM1 and TM2. (d) Relative reflectivity of the multilayer/antenna system compared with WinSpall simulation of a plain Ag layer, thickness 30 nm. The dashed curves represent the Lorentz fitting. (e) Transmission spectrum according to the incident angle (TM1 configuration). All data were recorded in air at $\lambda_{ex}$=650 nm.

The back bending and the broader reflectivity dip of the spectrum support the hypothesis of the antenna and multilayer mode hybridization in configuration TM1. The comparison of the reflectance curves of a multilayer/antenna system and a 30 nm-thick bare smooth Ag layer generated by an analytical model developed by using WinSpall simulation is shown in Figure 4d. To evaluate the resonance characteristics of the plain Ag thin film, a prism-based model was employed and the model was based on the coupling with a prism made of fused silica, which has a dielectric constant of $\varepsilon$ = 3.6, and an Ag layer, which has a dielectric constant of $\varepsilon$ = −15.2 + 0.4*$i$ [45]. As the results in Figure 4c, the minimum reflectivity of TM2 had shifted to a higher incident angle than the reference minimum. Compared to plain Ag, the multilayer/antenna system had a broader Full Width at Half Maximum of the Lorentzian fitting (FWHM=2.2 ± 0.2 deg versus FWHM=5.5 ± 2.1 deg). The current result validates the existence of two different modes that are typical of this system and effectively interact in the given configuration TM1 [46] showing multilayer/antenna mode hybridization, as confirmed by the transmission signal's double peak (Figure 4e).

*2.2. Emission as a fingerprint of molecular pattern*

The enhancement of the electromagnetic field intensity at the antennae is expected to influence the pattern of the emitted signal in the presence of the matter and the molecules [47]. To verify this hypothesis and the consequent impact on the optical sensing of the hybrid multilayer, experiments aimed at studying the optomechanical behavior were carried out. The fundamental vibrational frequency ($f$) of this system can be approximately calculated using the formula $f = v/d$. In this formula, $v$ represents the speed of sound propagation in metal ($v = 3 \times 10^3$ m s$^{-1}$) and $d$ denotes the typical dimension of the antenna where the propagation of the light is of greater significance. The value of $f$ has been evaluated to be approximately $f = 100\ MHz$. For the experiments, a mechanical mode was boosted by injecting an optical power of 5.0 mW into the system at an angle of incidence $\vartheta_{res} = 43.48\ deg$ and arranged in a prism coupling configuration [48]. The experimentation was conducted with a monochromatic beam which was necessary for system excitation and for measuring the system under the surface plasmon resonance conditions. The experimental findings supported the theoretical estimation, indeed, at the cavity resonance ($\vartheta_{res} = 43.48\ deg$), the FFT spectrum in Figure 5a highlights three fundamental peaks: a big peak at $\omega_c = 96\ MHz$, and two side peaks at $(\omega < \omega_c) = 95.75\ MHz$ and at $(\omega > \omega_c) = 96.25\ MHz$. The off-resonance spectra denote that the intensity of the side peaks, particularly that at $(\omega - \Delta\omega)$, varies with the angle of incidence, and it reaches its highest value at the incident angle of the resonance (Table 1). According to the definition of the decay rate estimated as the linewidth of the peak of the Lorentzian function that fits the experimental curve, the decay is equal to $\kappa = 1.1 \times 10^{-2}$ MHz and this showcases a high quality factor of $Q = 9.6 \times 10^3$. Therefore, by avoiding significant losses, the hybrid multilayer is characterized by having a high quality factor that might play a crucial role in enabling a strong coupling between the hybrid multilayer and the matter and therefore support an enhancement of the sensing ability [49]. It is also interesting to observe that it is unlikely that the peak behavior is caused by a wave-vector effect because there was only a small frequency shift. In turn, it is more probable that the emission spectrum is affected by the inelastic scattering of the vibration localized at the micrometric features. Moreover, the present inelastic scattering phenomenon is similar to the coupling between light scattering and acoustic phonon. Notably, the acoustic phonon, which denotes the mechanical mode stimulated within the multilayer by the laser pump, is amenable to investigation by means of Brillouin spectroscopy [48]. Under resonance conditions, the side peaks tend to significantly diminish, thereby emphasizing the impact of optical resonance on transmitted spectrum. The measurement was conducted in an aqueous environment by utilizing a light beam to boost the emitted mode at the resonance angle ($\vartheta_{h2o,res} = 46.5\ deg$). The aim of this experiment was to determine the impact of the surrounding medium on the resonance angle by measuring the mode of emission in an aqueous environment. Figure 5b illustrates that as the refractive index n decreases, the fundamental side peak frequency shift. Overall, the multilayer/SPP system's spectrum displays a low-frequency mode attributed to an acoustic vibration localized at the column features. Therefore, the localized vibrations of the cavity's antenna are conditions necessary for peak existence, however, the enhancement is attributed to the dipolar plasmon's resonant optical excitation, which specifically enhances the scattering from the vibrations localized on the feature.

To assess the sensing abilities, molecular sensitivity was investigated. To this aim, the surface of the hybrid multilayer was used to adsorb Congo red, which is classified as an azo dye. A solution containing a sample with a concentration of 1 mM was deposited at the inlet of the groove to facilitate capillary filling. The spectra of transmission were then recorded and shown in Figure 6a and Figure 6b.

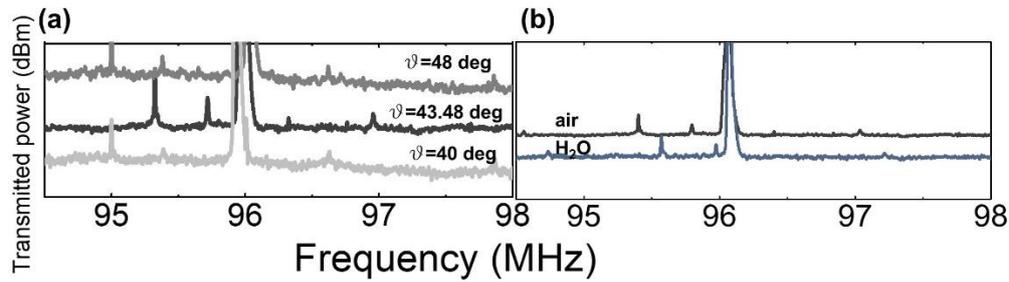

**Figure 5.** Transmitted power. (a) FTT spectrum of the optomechanical hybrid system at several angles of the angle of incidence. (b) FTT spectrum of the optomechanical hybrid system at $\vartheta_{res} = 43.48\ deg$ in an aqueous environment ($\vartheta_{h2o,res} = 46.5\ deg$).

The patterns recorded at various angles of incidence indicated a shift of the side modes with respect to the plain hybrid multilayer. In turn, when the concentration of the red dye was increased from $10^{-3}$ mM to 1 mM, the effect of inelastic scattering caused by the molecules was observed. Figure 6c shows that the intensity of the hybrid mode of the cavity also increases as the Congo red surface coverage factor associated with the concentration of the molecules in solution increases. Therefore, the localized vibrations of the micrometric features of the cavity are conditions necessary for peak existence, but the enhancement is attributable to the interaction between the hybrid multilayer and the molecular modes. However, it is remarkable that the differences between the last two concentrations (e.g. $10^{-2}$ mM to 1 mM) are negligible due to the saturation of the surface and the degradation of the *Q*-factor [49].

On the other side, the influece of the quality factor is visible when the laser wavelength is reduced down to $\lambda_{ex}$=520 nm and compared with the same pattern recorded at $\lambda_{ex}$=650 nm (Figure 6d). The dependence of the mode on the frequency of the excitation source suggests that the relative amplitude of the scattering from different modes can change because of the variation in the electronic-molecular vibrational mode coupling and entwines that a mechanical mode can be boosted by the laser. The relative amplitude of the scattering from different modes can change because of the variation in the electronic-vibrational coupling, as the laser wavelength is changed. Considering the vibrational modes as related to the strength of the electron-phonon coupling, the observation made above implies that the resonant optical excitation of a dipolar plasmon selectively enhances the scattering from acoustic vibrations localized on the features of the metal surface. Moreover, the diagram in Figure 6 shows a slight change in the optomechanical frequency recorded at the lower wavelength and underlines that for the system in question the effect of the cavity resonance wavelength shift is balanced by the variation of the quality factor. Indeed, because the optical behavior of the system depends on both the laser/multilayer detuning and the *Q*-factor value, the high *Q* of the hybrid multilayer permits that the shift in the cavity resonance wavelength has a negligible effect on the overall frequency and enables to maintain a stable frequency.

**Table 1.** Characteristics of the peaks of the hybrid multilayer/antenna system.

| $\vartheta_{inc}$ deg | $\omega < \omega_c$, Transmitted power, dB | FWHM, MHz | $\omega > \omega_c$, Transmitted power, dB | FWHM, MHz |
|---|---|---|---|---|
| 40 | 0.003 | 0.001 | 0.002 | 0.002 |
| 43.48 | 0.015 | 0.011 | 0.009 | 0.007 |
| 48 | 0.009 | 0.007 | 0.004 | 0.001 |

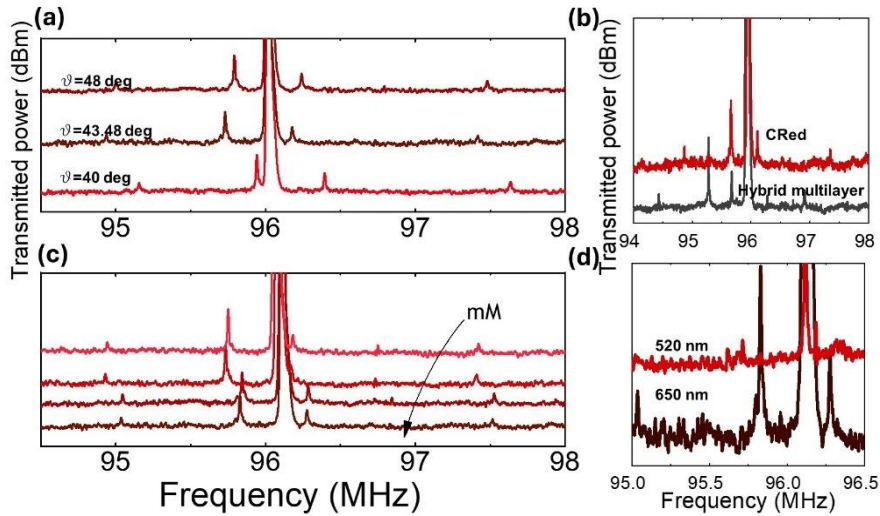

**Figure 6.** Sensing a dye. (a) Optomechanical spectrum of the cavity/SPP hybrid system following the adsorption of the Congo red at different angles of incidence. (b) The shift of the dye/cavity system relative to the plain cavity at $\vartheta_{res} = 43.48\ deg$. (c) Optomechanical spectrum of the cavity/SPP hybrid system following the adsorption of the Congo red at different concentrations. Concentration from top to bottom: $10^{-3}$ mM; $5\times10^{-3}$ mM; $10^{-2}$ mM; 1 mM. (d) Comparison of two excitation wavelengths, $\lambda_{ex}$=650 nm and $\lambda_{ex}$=520 nm (Congo red 1 mM).

## 2.3. The proof-of-concept design for protein detection

The transmitted mode's dependence on the frequency of the excitation source suggests that the scattering amplitude from different modes may vary due to changes in the coupling between electronic-molecular vibrations, allowing for molecular sensing. To demonstrate the sensitivity of the experiment, it was necessary to test a non-specific protein, the Bovine Serum Albumin (BSA). The spectrum of a solution containing 0.001 mg mL$^{-1}$ BSA in Phosphate Buffered Saline (PBS) was acquired and analyzed [50]. The sample was diluted using the procedure from the experimental section, and 1 μL of the protein solution was dropped into the groove to be measured. Figure 7a displays the protein transmitted spectrum compared to the naïve PBS reference.

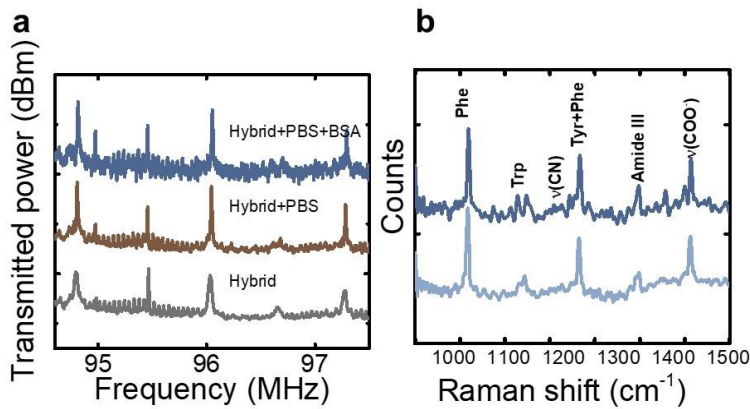

**Figure 7.** BSA characterization. (a) spectrum of hybrid multilayer following the adsorption of the PBS and BSA. (b) SERS as an average spectrum from CCD inspection at a low (bottom curve 10$^{-6}$ BSA mg mL$^{-1}$ in PBS) and a high (top curve 10$^{-3}$ BSA mg mL$^{-1}$ in PBS) concentration solution of BSA.

The introduction of samples causes the side peaks at $\omega > \omega_c$ to be modified, indicating that the samples inflict an alteration at the optical properties of the multilayer [51]. The surface-enhanced Raman spectroscopic signal was used to quantify the positive impact of amplifying the transmitted signal on molecular analysis and BSA sensing. A charge-coupled device (CCD) camera was used to record the emitted signal, and the data were extrapolated from the pictures using the technique outlined in reference [52] for producing a Raman spectrogram of the BSA. According to the method mentioned above, a low (10$^{-6}$ BSA mg mL$^{-1}$ in PBS) and a high (10$^{-3}$ BSA mg mL$^{-1}$ in PBS) concentration solution of BSA were tested; Figure 7b displays the average surface-enhanced Raman scattering spectra from 750 frames that were used to generate the heat map. The spectra of the two solutions display three distinct bands that are centered at 1001, 1159, and 1550 cm$^{-1}$. These bands correspond to the ring-breaking band phenylalanine residues (Phe), the C−N stretching vibration of phenylalanine and acid amides, and the tyrosine-phenylalanine residues. The visibility of peaks in a spectrum is largely dependent on the concentration of the sample. Specifically, higher concentrations result in a more pronounced and discernible peak spectrum, as documented in reference [53]. However, regardless of the concentration, a comparison between the on-resonance hybrid system and a standard SERS spectrum highlights a greater sensitivity in the former, indicating the possibility of capturing a more complex and complete Raman transition pattern.

At this stage, it is essential to draw attention to the advantages and limitations of the molecular analysis method introduced in this study. A thorough understanding of the method's strengths and weaknesses is critical to its successful implementation in real-world applications. In fact, by carefully weighing the benefits and drawbacks of this approach, researchers can make informed decisions about how best to leverage its potential for advancing scientific knowledge and improving outcomes in a variety of disciplines. Several techniques allow a precise and sensitive molecular readout. Table 2 underlines the advantages and disadvantages of the presented approach when compared with similar analytical methods with different strengths and limitations. It is interesting to observe that the limitation of the optomechanical detection relative to the nonlinear vibrational instabilities can be overcome by building up the significant vibrational state populations, required for nonlinear Raman, by controlling the light source properties (e.g. an intense short-pulse excitation).

**Table 2.** Comparison among molecular sensors.

| Analytical method | Sensor principle | Strengths | Limitations |
|---|---|---|---|
| Molecular vibration (this work) | Optical absorption and reflection | High sensitivity Flexibility Simple fabrication Potential anti-Stokes regime detection[54] | ps order of phonon lifetimes in molecules |
| Near-infrared (NIR)spectroscopy including Raman and SERS | Optical absorption and reflection | High sensitivity Fast readout Single component detection | Difficult calibration Limited quantification capability |
| Terahertz | Electromagnetic radiation | High penetrability in various materials Low photon energy | Low S/N Attenuation in water medium |
| X-Ray | Electromagnetic radiation | High resolution | High cost |

## 3. Conclusions

It has been demonstrated a hybrid multilayer to explore molecular spectroscopy. The layout consisted of metal and dielectric alternating layers with a thin film of Ag coated by plasmonic antennae. The complex optical architecture results in SPP/multilayer mode coupling demonstrated through the analysis of the reflectivity and the electromagnetic field distribution. The hybrid mode has been hypothesized to boost the sensitivity of the system and a transmitted signal. To investigate the coupling and gauge the strength, the multilayer was first adsorbed with a red dye, Congo red. The transmitted signal of the multilayer including the dye has been investigated at different wavelength and concentrations; then the BSA has been tested. To determine how improving the emitted signal affects the analysis of molecules and the detection of BSA, a Raman signal was measured. The results showed that the proposed method is more sensitive, which allows for a more detailed pattern of the spectrum with more Raman transitions to be captured.

## 4. Experimental Experimental section

### 4.1. Fabrication of the multilayer and characterization

A multilayer consisting of noble metals and dielectric was used for the experiments. The detailed scheme of the multilayer can be found in Table 3 and schematized in Figure 1a. The multilayer features a purposeful groove to securely hold the sample in place. The groove was created through a series of etching/passivation cycles, ultimately resulting in an etching depth of approximately 225 µm. Following the process, the surface of the multilayer displayed a micrometric roughness.

**Table 3.** Refractive index and geometry of the layers in multilayer and micrometric features.

| multilayer | | Micrometric feature | | Refractive index |
|---|---|---|---|---|
| layer ID | height | parameter | size | $n_i$ @ $\lambda$=650 nm |
| Pt | 500 nm | gap | ~1 µm | 2.38+4.26i |
| SiO$_2$ backside | 500 nm | diameter | 1.4 µm | 1.44 |
| Si | 30 µm | Si height | 30 µm | 3.85+0.018 i |
| SiO$_2$ frontside | 1 µm | SiO$_2$ height | 1 µm | 0.052+4.41i |
| Ag | 30 nm | Ag height | 30 nm | |

In conducting optical measurements, I utilized the Kretschmann configuration approach [55–57]. The system was stimulated using a tunable continuous-wave laser featuring a 650 nm wavelength through a beam splitter. Prior to each measurement, the system was allowed ample time to stabilize. To ensure accurate measurements, I utilized a Thorlabs RC12FC-P01 for laser collimation and a double Glan-Taylor Calcite Polarizer for p-polarization. Additionally, I minimized noise using an optical iris diaphragm from Thorlabs (D25SZ). After being filtered and collimated through a Glan Thomson polarizer from Thorlabs that operates within the 650-1050 nm range, the reflected beam was collected by a Si photodiode with a bandwidth of 960 nm to capture the signal. To minimize any noise, a ceramic disk capacitor was also employed. The Raman module was affixed to the front of the prism, and a silicon photodiode with a wavelength range of 340-1100 nm was utilized to collect the light. Likewise, a ceramic disk capacitor was utilized to reduce noise. A PicoTech series 5000 spectrum analyzer was connected to the photodiode, and a CCD was utilized to detect the emitted signal for Raman analysis. Measurements of SERS spectra were conducted using a microspectroscopic setup featuring a single-stage spectrograph equipped with a liquid-nitrogen-cooled charge-coupled device detector. Raman scattering was excited through a 60× immersion objective (Olympus, Germany) in 180° backscattering geometry, utilizing a laser intensity of $5.7 \times 10^5$ W cm$^{-2}$ and a 785 nm wavelength. Spectra were obtained from measurements on droplets of approximately 1 pL to ensure a high degree of accuracy and precision in spectral data collection. The protein concentration was chosen in the range of intracellular protein concentration values (0.02-0.55 g mL$^{-1}$). The BSA was prepared by dissolving the Sigma-Aldrich-purchased BSA in Milli-Q water. Subsequently, 70 µL of the resulting solution was carefully transferred onto a calcium fluoride plate, where Raman spectra were obtained from the droplet. The acquisition time for the spectra was set to 10 seconds.

*4.2. Numerical model*

The simulation of the theoretical model was conducted using Matlab (Ed. 2023). The script implemented the Rigorous Coupled Wave Analysis (RCWA) in the frequency domain, following the model presented in reference [58]. The analysis computed the diffracted amplitudes and diffraction efficiencies of finite-size structures made up of the stack layer.

RCWA relied on the computation of the eigenmodes in all layers of the structure with the micrometric features based on a Fourier basis [59] and on a scattering matrix approach, for recursively relating the mode amplitudes in the different layers. The sum of the reflected and transmitted efficiency and the loss per layer divided by half period were used to determine the absolute error ($<10^{-5}$). I sourced the refractive index for each layer from specific references. The refractive index for fused silica was obtained from the WVASE library, while Palik's data [60] was used for Ag, Pt, Si, and $SiO_2$. Schott's provided the refractive index for NBK7 glass. The analysis utilized the wavelength and angle of incidence angle as the sweeping parameters.

**CONFLICT OF INTEREST**：  The author declares no conflict of interest.